\documentclass[12pt,english]{article}
\usepackage[T1]{fontenc}
\usepackage[latin9]{inputenc}
\usepackage{geometry}
\usepackage{amstext}
\usepackage{amssymb}
\usepackage{graphicx}
\usepackage[numbers,sort&compress]{natbib}
\usepackage{babel}
\begin{document}
\title{Residual Smoothing: Using Mocks to Correct Model Covariance Matrices}
\author{Ross O'Connell}
\maketitle
\begin{abstract}
Covariance matrix estimation is a challenging problem in cosmology.
Recent work has shown that model covariance matrices can be precise,
and that at relatively large scales they can also be accurate. We
introduce a data-driven method that can identify features from a mock
covariance matrix that are missing from a corresponding model, then
incorporate them into the model without significantly degrading the
model's precision. We apply this method to a BOSS-like survey and
extend a model covariance to be valid at scales relevant for measurements
of Redshift Space Distortions ($8-40\,h^{-1}\text{Mpc}$), where the
galaxy field is significantly non-Gaussian.
\end{abstract}

\section{Introduction}

Covariance matrix estimation is a necessary part of many cosmological
analyses. It is a challenging problem because we have only one sky
to observe, and so cannot directly generate multiple sets of independent
observations. See \citep{Taylor:2014ota,Dodelson:2013uaa,Percival:2013sga}
for foundational work explicating the covariance matrix problem, and
\citep{Zhu:2014ica,Padmanabhan:2015vlf,Escoffier:2016qnf,Joachimi:2016xhk,Klypin:2017,Howlett:2017vwp,Zhu:2018edv,Barreira:2018jgd,Wang:2019sru,Philcox:2019ued,Philcox:2019xzt}
for some of the recent work that addresses the problem.

Broadly speaking, there are two ways to solve this problem. The first
is to generate reasonably accurate numerical simulations of the evolution
of the universe, or ``mocks'', and perform sample statistics with
the mocks. The resulting covariance matrix is often assumed to be
accurate, but suffers from challenging trade-offs between the number
of mocks generated (with each being computationally expensive) and
the level of noise on the final covariance matrix. The other approach
is to use theoretical insights to produce an analytic or semi-analytic
model covariance matrix. These approaches generally feature high precision
at minimal computational expense, but the accuracy of the resulting
covariance matrix rests on the underlying theoretical model, which
may be incomplete or inaccurate.

In this note we introduce a simple method to reconcile the numerical
and theoretical approaches. We assume that both mocks and a model
covariance matrix are available, focus our attention on the residual
difference between the two, then smooth that residual. We then add
the smoothed residual back to the original model covariance matrix
to achieve a result that combines the accuracy of the mocks with the
precision of the model. Our improvements rest on the validity of the
smoothing technique; we use the Kullback-Leibler divergence, a robust
tool from information theory, to track the quality of the resulting
covariance matrix, and cross-validation to limit the degree of overfitting/undersmoothing.

We demonstrate this approach on a BOSS-like galaxy survey. In \citep{OConnell:2015src}
we developed a model covariance matrix that is largely Gaussian, and
incorporates short-scale non-Gaussianity through a shot-noise rescaling
parameter. This model was found to be accurate at separations relevant
for measurements of baryon acoustic oscillations (BAO) ($\gtrsim80\,h^{-1}\text{Mpc}$),
but at separations relevant for measurements of redshift space distortions
(RSD) ($\lesssim40\,h^{-1}\text{Mpc}$) its accuracy was compromised,
presumably because it did not incorporate mid-scale non-Gaussianity
of the galaxy field. Applying residual smoothing we are able to correct
the model and restore these non-Gaussian contributions.

In Section \ref{sec:Projection} we introduce the family of projection
operators that are used to smooth the residual. If the correlation
function is estimated in $n$ bins, then this family is $2^{n}$-dimensional.
In Section \ref{sec:Algorithm} we develop an algorithm to identify
relevant members of this family and assess their performance. In Section
\ref{sec:Application} we apply the method to a BOSS-like survey and
see that it identifies necessary corrections to the model covariance
matrix with minimal overfitting. We conclude in Section \ref{sec:Discussion}.

\section{\label{sec:Projection}Projection Operators and Smoothing}

Suppose that we have a model covariance matrix, $C_{\text{model}}$,
and a covariance matrix computed from a modest number of mocks, $C_{\text{sample}}$.
We define the residual between these covariance matrices as 
\begin{equation}
\Delta=C_{\text{sample}}-C_{\text{model}}\,.
\end{equation}
The residual $\Delta$ can non-zero both because of noise in $C_{\text{sample}}$,
and also because of bias \footnote{Throughout this note we will assume that $C_{\text{sample}}$ provides
an unbiased estimate of the true covariance matrix. In practice this
may not be true, but we believe the problem of correcting a model
covariance matrix will be largely separate from the problem of constructing
more accurate mocks.} in $C_{\text{model}}$ relative to $C_{\text{sample}}$. Our goal
is to generate $\tilde{\Delta}$, a smoothed version of $\Delta$,
which minimizes the contributions from noise and isolates the bias
of $C_{\text{model}}$ relative to $C_{\text{sample}}$. We can then
use $C_{\text{model}}+\tilde{\Delta}$ as an improved covariance matrix
estimate, with less bias than $C_{\text{model}}$ alone and less noise
than $C_{\text{sample}}$.

We will use projection operators to smooth $\Delta$. Mathematically,
a projection operator $\Pi$ is idempotent, 
\[
\Pi^{2}=\Pi\,.
\]
The intuition is that the operator restricts to a subspace, and that
subsequent applications of the projection operator do not produce
any further change. In our case the projection operators will be matrices
of the same shape as the covariance matrix, and will act on the residual
as 
\begin{equation}
\tilde{\Delta}=\Pi\Delta\Pi^{T}\,.
\end{equation}
We then combine the smoothed residual $\tilde{\Delta}$ with the original
model to yield the corrected model, 
\begin{equation}
\tilde{C}=C_{\text{model}}+\tilde{\Delta}\,.
\end{equation}
In order for projection to lead to smoothing, we must be able to identify
a subspace that contains the bulk of the bias in $C_{\text{model}}$
while excluding most of the noise from $C_{\text{sample}}$.

The family of projection operators that we consider are built from
eigenvector decompositions of matrices. Suppose that an $n\times n$
symmetric matrix $C$ is decomposed as 
\begin{equation}
C=Q\times\text{diag}\left(\lambda_{1},\lambda_{2},\dots,\lambda_{n}\right)\times Q^{T}\,,
\end{equation}
where the $i$th column of $Q$ is the $i$th eigenvector of $C$,
and $\lambda_{i}$ is the $i$th eigenvalue of $C$. We can construct
a large family of projection operators as 
\begin{equation}
\Pi=Q\times\text{diag}\left(\delta_{1},\delta_{2},\dots,\delta_{n}\right)\times Q^{T}\,,
\end{equation}
where $\delta_{i}$ is either one or zero. A familiar example of this
approach would be to set $\delta_{i}=1$ for the $k$ largest eigenvalues
and $\delta_{i}=0$ otherwise. The resulting projection operator would
pick out the same $k$-dimensional subspace that would emerge from
principal component analysis (PCA). 

We have three candidates sources for the eigenvectors $Q$: $C_{\text{sample}}$,
$C_{\text{model}}$, and $\Delta$. Our goal is to cleanly separate
bias from noise, which is difficult to do if the eigenvectors themselves
have a significant amount of noise. For that reason we will use the
eigenvectors of $C_{\text{model}}$. While $C_{\text{model}}$ and
$\Delta$ are qualitatively different from one another, they share
many important features pertaining to bin ordering, scaling, etc.,
and so the eigenvectors from $C_{\text{model}}$ can efficiently capture
the features\footnote{We did experiment with a simpler PCA of $\Delta$, but noise on the
eigenvectors made the results underwhelming.} of $\Delta$.

After choosing the eigenvectors $Q$, we need to determine which $\delta_{i}$
should be non-zero. Note that because $C_{\text{model}}$ and $\Delta$
are quite different matrices, we do not expect (and in the following
do not find) that the leading eigenvectors are the ones that should
be included in the final subspace (i.e. have $\delta_{i}=1$). Rather
we must search among the $2^{n}$ possible projection operators to
identify the one that provides the optimal projection.

\section{\label{sec:Algorithm}Learning the Projection Operator}

In this section we describe in detail our method for identify the
preferred projection operator. Broadly speaking, there are two parts
to this procedure. First, we repeat the following many times:
\begin{enumerate}
\item Randomly split the available mocks into equal training and test sets,
compute sample covariance matrices $C_{\text{training}}$ and $C_{\text{test}}$.
\item Invert $C_{\text{test}}$ and apply the usual Wishart correction to
find $\Psi_{\text{test}}$.
\item Compute the residual $\Delta=C_{\text{training}}-C_{\text{model}}$.
\item Apply the Metropolis-Hastings algorithm to the $\delta_{i}$ to find
a projection operator that minimizes the Kullback-Leibler (KL) divergence,
$\text{KL}\left(\tilde{C},\Psi_{\text{test}}\right)$. The KL divergence
is described below in Section \ref{subsec:KLDivergence}.
\end{enumerate}
The repeated splitting of mocks into training and test sets is known
as Monte Carlo cross-validation. For any particular split we will
find a large number of modes included in the optimal projection operator.
By iterating over splits we can better identify which modes consistently
appear, and which modes appear infrequently and are associated with
noise.

Because our final projection operator must have only 0's and 1's for
the $\delta_{i}$, we cannot simply average the results of each run
to find a consensus projection operator. Instead we track how many
times each mode was included in an optimal projection operator, producing
a ranking of the modes. In the final projection operator modes are
included according to that ranking. A second round of cross-validation,
described in Section \ref{subsec:XV2}, determines how many of those
modes to include.

We use Monte Carlo cross-validation in order to reduce overfitting
of the projection operator. As we are comparing statistically independent
training and test sets, the projection operator should not be able
to adapt to noisy features in the training set. However, we assume
a limited number of mocks are available, and so noisy features in
the entire set of mocks will be split between the training and test
sets for each iteration of Monte Carlo cross-validation. This opens
the door to a limited amount of overfitting, and we will see in Section
\ref{sec:Application} that it has a modest impact on our results.

\subsection{\label{subsec:KLDivergence}The Kullback-Leibler Divergence}

To quantify the difference between a proposed covariance matrix and
a sample covariance matrix, we use the Kullback-Leibler (KL) divergence
\citep{Kullback:1951aa}. The KL divergence is a tool from information
theory that provides a natural measure of the dissimilarity between
two distributions. For two multivariate-normal distributions with
the same mean, it takes a simple form,
\begin{equation}
\text{KL}\left(C_{1},\Psi_{2}\right)=\frac{1}{2}\left[\text{tr}\left(C_{1}\Psi_{2}\right)-\log\left(\det\left(C_{1}\right)\right)-\text{log}\left(\text{det}\left(\Psi_{2}\right)\right)-n\right],
\end{equation}
where $C_{1}$ is the $n\times n$ covariance matrix of the first
distribution and $\Psi_{2}=C_{2}^{-1}$ is the precision matrix of
the second distribution. When $C_{1}$ and $\Psi_{2}$ are known exactly,
larger values of the KL divergence correspond to greater differences
between the two matrices. If, on the other hand, $C_{1}$ is a sample
covariance computed from $n_{s}$ samples whose true covariance matrix
is $C_{2}$, then $\text{KL}\left(C_{1},\Psi_{2}\right)\sim n^{2}/4n_{s}$,
so noise also increases the KL divergence.

\subsection{\label{subsec:XV2}How many modes?}

As described above, the first round of cross-validation tells us how
often each mode is included in an optimal projection operator. At
this we could apply an arbitrary cut on frequency, perhaps after looking
at the results of the first round of cross-validation, and have a
useful projection operator. Here we develop a more systematic approach.
Our goal is to automatically determine the optimal number of modes
to include, and thus the optimal projection operator.

For the second round of cross-validation we repeat the following many
times:
\begin{enumerate}
\item Randomly split the available mocks into equal training and test sets,
compute sample covariance matrices $C_{\text{training}}$ and $C_{\text{test}}$.
\item Invert $C_{\text{test}}$ and apply the usual Wishart correction to
find $\Psi_{\text{test}}$.
\item Compute the residual $\Delta=C_{\text{training}}-C_{\text{model}}$.
\item For each $k$ of interest, construct a projection operator $\Pi_{k}$
with $\delta_{i}=1$ for the $k$ highest-ranked modes. Use it to
generate a corrected model $\tilde{C}_{k}$.
\item Compute $\text{KL}\left(\tilde{C}_{k},\Psi_{\text{test}}\right)$
for each $k$ of interest.
\end{enumerate}
Having found the distribution of KL divergences for each value of
$k$, it is then straightforward to determine which $k$ leads to
the smallest KL divergence, and is therefore optimal.

Our general expectation is that the first few modes will bring $\tilde{C}$
closer to $\mathbf{E}\left[C_{\text{sample}}\right]$, reducing the
KL divergence. As we add more modes the smoothed residual $\tilde{\Delta}$
incorporates more of the noise of $C_{\text{test}}$, and so we expect
the KL divergence to eventually increase. Competition between these
two effects would then produce an isolated minimum in the KL divergence
as a function of $k$. In cases where $C_{\text{model}}$ is already
very close to $\mathbf{E}\left[C_{\text{sample}}\right]$ it is possible
that this minimum will happen at $k=0$, indicating that we should
simply use the uncorrected $C_{\text{model}}$.

\section{\label{sec:Application}Application to a BOSS-like survey}

To demonstrate the efficacy of this approach we use 1,000 QPM mocks
\citep{White:2013psd}, generated for the analysis of the NGC portion
of the BOSS survey \citep{2013AJ....145...10D}, and a model covariance
matrix generated using Rascal \citep{OConnell:2015src} for the same
survey. We will consider a correlation function estimated in $r-\mu$
coordinates, estimated in bins of width $\Delta r=4\,h^{-1}\text{Mpc}$
and $\Delta\mu=0.1$. The model covariance includes a shot-noise rescaling
parameter $a=1.06$, in agreement with previous work \citep{OConnell:2018oqr}.

We consider two ranges of $r$-values. The high-$r$ range encompasses
100 bins over $r=96-136\,h^{-1}\text{Mpc}$. This is a range of separations
where we expect the galaxy field to be very Gaussian, and is relevant
for measurements of baryon acoustic oscillations (BAO). In previous
work we have found the model covariance matrix to be quite accurate
\citep{OConnell:2015src}, and so we anticipate a smoothed residual
that is small or zero. The low-$r$ range includes 100 bins over $r=8-48\,h^{-1}\text{Mpc}$.
This range of scales is relevant for measurements of redshift space
distortions (RSD). It is also a range of scales where the galaxy field
is significantly non-Gaussian, and consequently we expect that the
model covariance matrix will be biased. The smoothed residual should
reduce or eliminate this bias, extending the model covariance matrix
to a previously inaccessible range of $r$.

\begin{figure}
\includegraphics{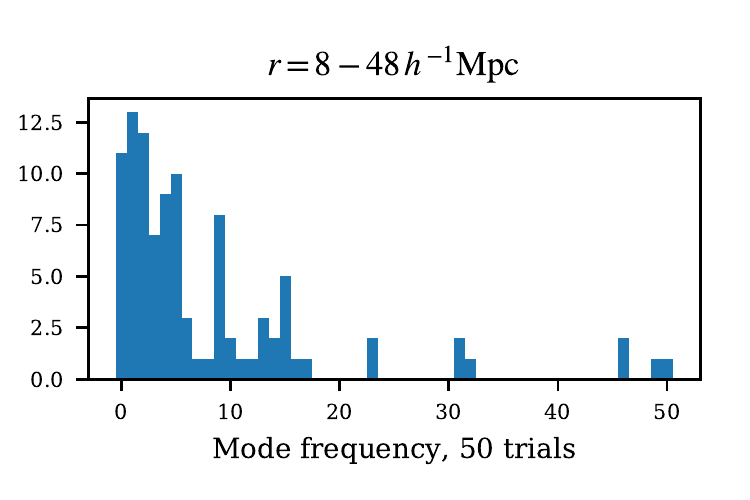}\hfill{}\includegraphics{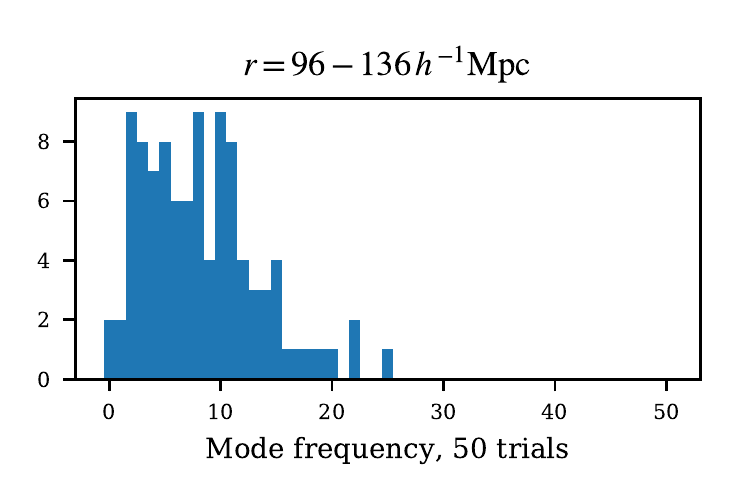}

\caption{\label{fig:ModeCounts}A histogram of mode frequencies for 50 repetitions
of Monte Carlo cross-validation. Modes that appear frequently correspond
to systematic differences between the model covariance matrix and
the true covariance matrix for the mocks. We see several such modes
for the low-$r$ range, but do not see them in the high-$r$ range.}
\end{figure}

We begin by using Monte Carlo cross-validation to identify relevant
modes, as described in Section \ref{sec:Algorithm}, repeating the
process 50 times. A histogram of mode frequencies is provided in Figure
\ref{fig:ModeCounts}. For the low-$r$ range we find several modes
that appear in all or almost all of those repetitions, providing strong
evidence that they should be used for residual smoothing. Such modes
do not appear in the high-$r$ range, suggesting that the model covariance
matrix is accurate and that the smoothed residual should be small
or zero.

\begin{figure}
\includegraphics{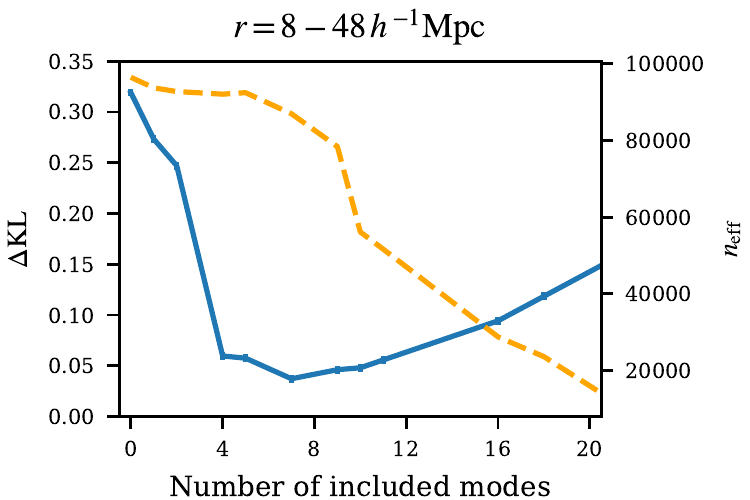}\hfill{}\includegraphics{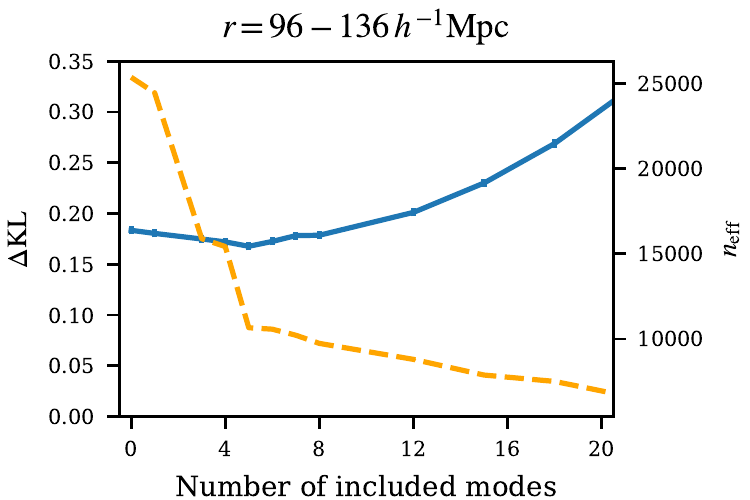}

\caption{\label{fig:KL_Graph}Improvement in the KL divergence between the
sample covariance (computed with mocks) and the model covariance (including
a smoothed residual) vs. the number of modes included in the projection
operator for the smoothed residual. The dashed line indicates the
noise level in terms of $n_{\text{eff}}$, an effective number of
mocks. In the low-$r$ range we expect a significant bias in the model
(without residual), and observe a dramatic improvement by adding a
modest number of modes. In the high-$r$ range the model (without
residual) performs well, and adding the smoothed residual leads to
negligible improvements. The increase in $\Delta\text{KL}$ as the
number of modes grows occurs because the projection operator admits
an increasing amount of noise to the smoothed residual.}
\end{figure}

We rank the modes by how frequently they appear, then add them in
that order to the projection operator. Each time we add a new mode,
we perform a new round of Monte Carlo cross-validation to determine
the KL divergence between the new model (with smoothed residual) and
the sample covariance. The results for the low-$r$ and high-$r$
ranges are shown in Figure \ref{fig:KL_Graph}. In order to better
illustrate the improvement in the KL divergence, we have subtracted
off the expected KL divergence between the model (without residual)
and the sample covariance of 500 draws with that model covariance.

For the low-$r$ range we observe a dramatic improvement in the KL
divergence using a projection operator with four modes, and modest
further improvement from including seven modes. This is a clear indication
that the smoothed residual brings the model closer to the true covariance
matrix for the mocks. Beyond that, adding additional modes \emph{increases}
the KL divergence, as the projection operator admits more noise to
the smoothed residual (or, equivalently, smooths less aggressively).
Adding the smoothed residual extends the validity of the model covariance
to low scales where it was previously unusable, and is the primary
result of this note.

For the high-$r$ range we observe only modest improvement in the
KL divergence. This is consistent with our expectation that the model
(without residual) is quite accurate at these scales. As the number
of modes increases, we see the same increase in the KL divergence
that we observed for the low-$r$ range.

As an auxiliary measure of the new model we also compute the effective
number of mocks, $n_{\text{eff}}$, as defined in \citep{OConnell:2015src}.
This quantifies the noise on the covariance matrix. The original model
is generated through numerical integration, so it is not noiseless.
As we add more modes to the smoothed residual, we expect the noise
to increase, and that consequently $n_{\text{eff}}$ should decrease.
$n_{\text{eff}}$ assumes that the noise on the covariance matrix
is Wishart, which is definitely not true for our model with or without
the smoothed residual, so we are less interested in exact numbers
and more interested in large changes. For the low-$r$ range Figure
\ref{fig:KL_Graph} shows that we can add 4-9 modes without significantly
increasing the level of noise in the covariance matrix. For the high-$r$
range we find that the noise increases significantly when any modes
are added.

\begin{figure}
\begin{centering}
\includegraphics{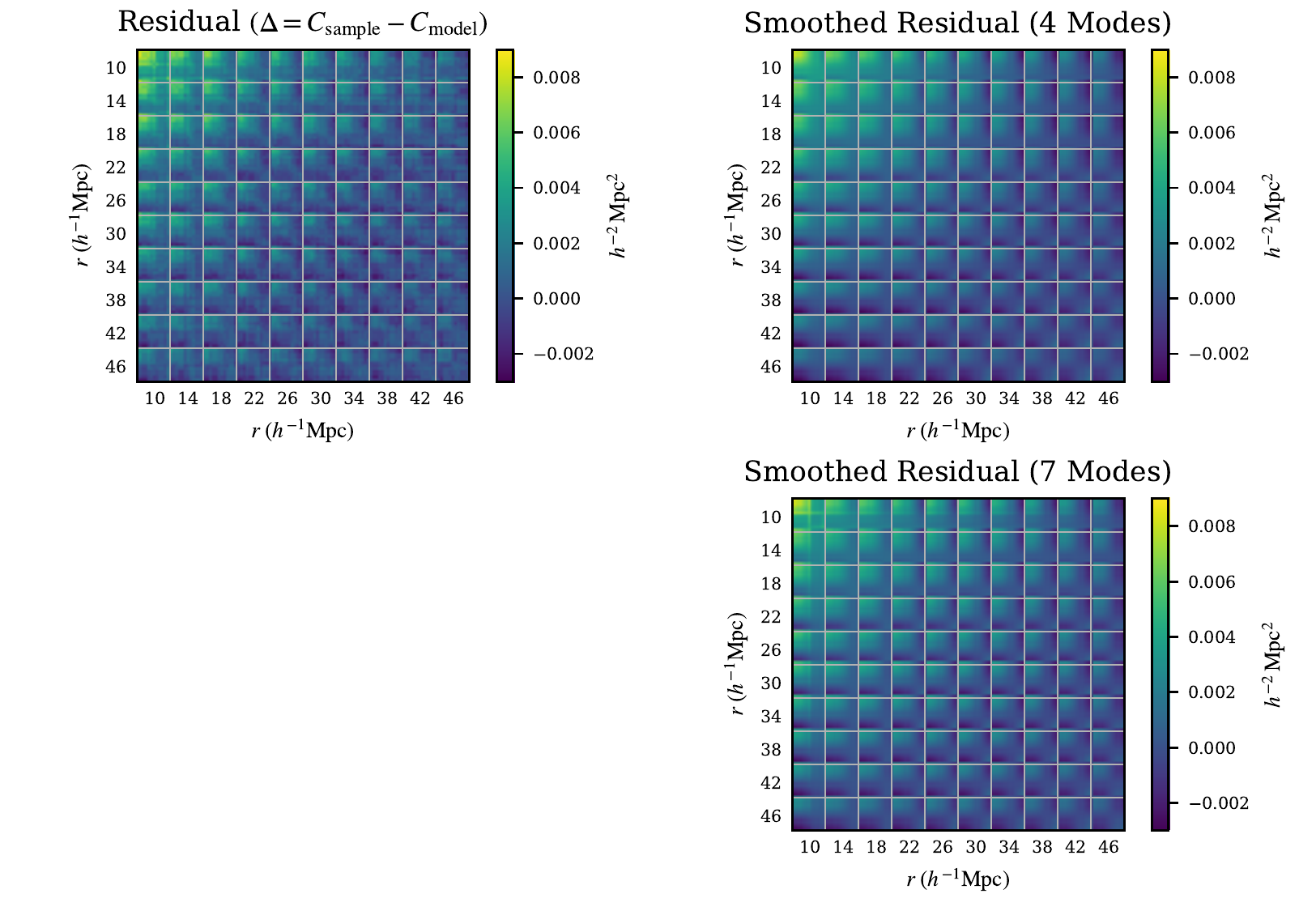}
\par\end{centering}
\caption{\label{fig:Residuals} The unsmoothed covariance matrix residual for
the low-$r$ range, along with two smoothed residuals using different
numbers of modes. Entries $C_{ab}$ are whitened by a factor of $r_{a}r_{b}$
to remove the leading $r$-dependence. Bins are grouped according
to their $r$ values, and within each group $\mu$ values increase
from $0.05$ to $0.95$. Both smoothed residuals are credible corrections
to the model covariance matrix, but the question of which is preferred
is difficult. Seven modes leads to the smallest KL divergence, but
sharp features in the $r=8-12\,h^{-1}\text{Mpc}$ are strongly reminiscent
of noise.}
\end{figure}

We would like to make a sharp statement about the optimal number of
modes to include in the projection operator in each case. However,
as discussed above, the Monte Carlo cross-validation we have undertaken
reduces, but does not eliminate, overfitting. In the low-$r$ case
the mode frequencies shown in Figure \ref{fig:ModeCounts} provide
strong evidence that four modes should be included in the projection
operator, with those modes appearing in almost every Monte Carlo run.
The next three modes appear a little more than half the time, so the
evidence in their favor is weaker. When we look at the impact of these
modes on the KL divergence in Figure \ref{fig:KL_Graph}, the bulk
of the improvement is dues to the initial four modes, but the KL divergence
does decrease when the next three modes are included. In Figure \ref{fig:Residuals}
we plot the actual residual and, for comparison, the smoothed residuals
with four modes and with seven modes. When four modes are included
the resulting residual is quite smooth. With seven modes a sharp feature
appears in the $r=8-12\,h^{-1}\text{Mpc}$ range, and it is plausible
that this feature is a result of noise, not signal. We believe that
the best interpretation is that the additional three modes constitute
overfitting, but a more rigorous statistical analysis would be required
to prove that this is the case.

\begin{figure}
\includegraphics{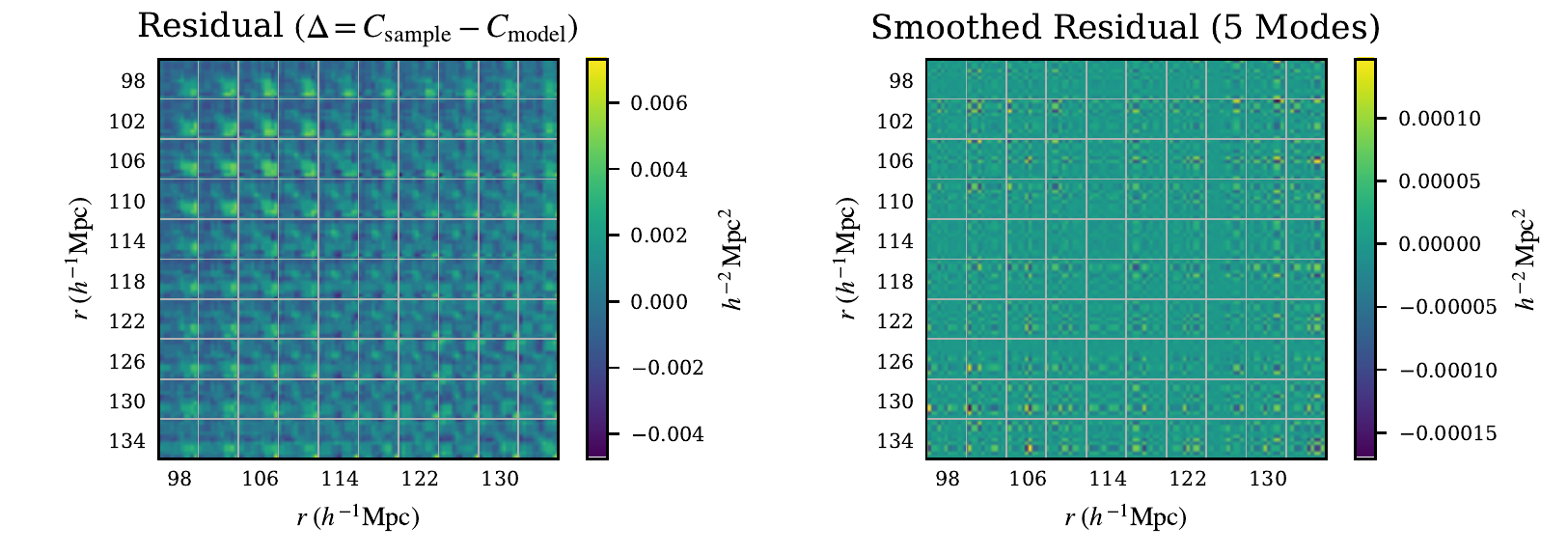}

\caption{\label{fig:BadResiduals}The unsmoothed covariance matrix residual
for the low-$r$ range, along with the smoothed residual using five
modes. Entries $C_{ab}$ are whitened by a factor of $r_{a}r_{b}$
to remove the leading $r$-dependence. Although the unsmoothed residual
appears to exhibit structure, it has not survived cross-validation
and we conclude that it is noise. The smoothed residual is close to
zero -- note the different scales for the two plots.}

\end{figure}

In the high-$r$ case the question is ``should we include a smoothed
residual at all?'' Figure \ref{fig:KL_Graph} shows a decrease in
the KL divergence with five modes included, but the decrease is quite
modest -- smaller than the decrease in the low-$r$ case when increasing
from four to seven modes. If we look at the mode frequencies in Figure
\ref{fig:ModeCounts}, we see that the five modes we might include
occurred less frequently than the three modes we found spurious in
the low-$r$ case. Finally, in Figure \ref{fig:BadResiduals} we plot
the actual residual and smoothed residual, and observe no structure
of note in the smoothed residual. We conclude that the drop in KL
divergence shown in Figure \ref{fig:KL_Graph} is a result of overfitting,
and that for high $r$ the model should be used without a smoothed
residual.

\section{\label{sec:Discussion}Discussion}

We have introduced a method for combining a model covariance matrix
with a set of mocks. The resulting covariance matrix can incorporate
features of the mocks that were not reflected in the model, while
maintaining low levels of noise. To demonstrate the method we constructed
a covariance matrix for a BOSS-like survey that would be applicable
at separations of $8-40\,h^{-1}\text{Mpc}$, scales where a Gaussian
model covariance matrix would not be valid. While this represents
valuable progress in covariance matrix estimation, we briefly discuss
two limitations of the method.

The smoothed residual approach rests on our ability to select eigenvectors
of the model covariance matrix which efficiently describe the difference
between the model and mock covariance matrices. As discussed in the
text, our method appears to select a few more modes than would be
optimal. Our use of cross-validation reduces this overfitting, but
does not eliminate it entirely. Further statistical work might reveal
an approach that corrects this.

A second limitation is that we assume that the mock covariance matrix
is not biased, relative to the true covariance matrix for the survey.
This is a strong assumption, and one that could be violated in a meaningful
way in practical applications. We recently demonstrated that models
of non-Gaussianity can be calibrated directly against the survey using
jackknife techniques \citep{OConnell:2018oqr}, but the approach introduced
here does not include a model for the residual and so cannot be connected
to the survey in this way. The accuracy of the mocks used is therefore
a limiting factor.

\bibliographystyle{hunsrtnat}
\bibliography{CovRefs}

\end{document}